\documentclass{article}
\usepackage{amssymb}
\usepackage{amsmath}
\usepackage{graphicx}
\usepackage{epsfig}

\newtheorem{theorem}{Theorem}
\newtheorem{acknowledgement}[theorem]{Acknowledgement}

\input{tcilatex}
\begin{document}

\title{Entanglement-assisted capacities of time-correlated amplitude-damping
channel}
\author{Nigum Arshed and A. H. Toor \\
\textit{Department of Physics, Quaid-i-Azam University}\\
\textit{Islamabad 45320, Pakistan}}
\maketitle

\begin{abstract}
We calculate the information capacities of a time-correlated
amplitude-damping channel, provided the sender and receiver share prior
entanglement. Our analytical results show that the noisy channel with zero
capacity can transmit information if it has finite memory. The capacities
increase as the memory increases attaining maximum value for perfect memory
channel.
\end{abstract}

\section{Introduction}

Entanglement, a purely quantum phenomena, describes global states for
composite systems which cannot be written as product of individual system
states. Once seen as a counterintuitive feature of quantum mechanics it has
emerged as a useful resource for quantum information and computation \cite%
{Horodecki 2009}. It performs tasks which cannot be accomplished by
classical resources such as, quantum cryptography \cite{Ekert 1991}, quantum
dense coding \cite{Bennett and Wiesner 1992}, quantum teleportation \cite%
{Bennett Brassard Crepeau Jozsa Peres and Wootters 1993} and communication
of reliable information over quantum channels.

Quantum channels model the noise processes which occur in quantum systems
due their unavoidable interaction with the environment \cite{Nielson and
Chuang}. There are broadly two types of quantum channels. Memoryless quantum
channels, for which the channel acts independently over each channel input
and quantum memory channels where the noise over successive channel uses
exhibits some correlation. In practice, memoryless channel is an
approximation as the channel properties are modified by its uses. The
maximum amount of information reliably transmitted over a channel, per
channel use is known as its capacity \cite{Thomas and Cover}. In comparison
to classical channels \cite{Shannon 1948}, the theory of quantum channel
capacities is more involved as more than one capacities are associated with
a quantum channel \cite{Bennett and Shor 2004}.

Early research in quantum channel capacities focused on memoryless channels
[9-24]. The studies of quantum memory channel capacities have attracted a
lot of attention lately and many interesting results were reported [25-36].
It was established that memory increases the capacity of a quantum channel
and beyond a certain memory threshold, classical capacity for both unital
and non-unital channels is higher for entangled state encoding [25-27].
Noisy quantum channels have non-zero quantum capacity in the presence of
memory as it suppresses the channel noise, similar to superactivation
phenomenon \cite{Smith and Yard 2008, Jahangir Arshed and Toor 2012}.
Entanglement-assisted classical capacity for unital quantum channels also
increases if noise over successive channel uses is correlated \cite{Arshed
and Toor 2006}.

In this work, we study information transmission over an amplitude-damping
channel. We assume that the noise over consecutive uses of the channel is
correlated and calculate its classical and quantum capacities. Entanglement,
unlimited or limited, is shared prior to the communication. Our results show
that the capacities of the channel increase as its memory increases. The
channel noise is suppressed by its memory and its capacity to transmit
information is always non-zero in the presence of memory.

The article is organized as follows. In Sec. 2, we review quantum channels
and give a brief description of entanglement-assisted capacities in Sec. 3.
In Sec. 4, we discuss the time-correlated amplitude-damping channel. We
calculate its entanglement-assisted capacities in Sec. 5. Finally, in Sec. 6
we discuss the results and present our conclusions.

\section{Quantum channels}

The basic task of quantum communication is to transmit information encoded
in quantum states across a quantum channel reliably \cite{Bennett and Shor
1998}. The reliability of communication depends upon the channel noise.
Mathematically, a quantum channel $\mathcal{N}$\ is a completely positive
and trace preserving map of a quantum system from an initial state $\rho
_{s} $ to the final state \cite{Nielson and Chuang},%
\begin{equation}
\mathcal{N\colon }\rho _{s}\rightarrow \mathcal{N}\left( \rho _{s}\right) .
\label{CPTP map}
\end{equation}%
It is assumed that the total system $\mathcal{H}_{s}\otimes \mathcal{H}_{e}$%
, consisting of the quantum system $\rho _{s}$ and its environment $\rho
_{e} $, is closed and undergoes unitary evolution. After interaction final
state of the system is given by%
\begin{equation}
\mathcal{N}\left( \rho _{s}\right) =\text{Tr}_{e}\left[ U\left( \rho
_{s}\otimes \rho _{e}\right) U^{\dagger }\right] ,  \label{Unitary evolution}
\end{equation}%
where Tr$_{e}$ is partial trace over the environment. This can be described
by Kraus operators acting on system $\rho _{s}$ \cite{Kraus 1983},\ as%
\begin{equation}
\mathcal{N}\left( \rho _{s}\right) =\sum_{i}A_{i}\rho _{s}A_{i}^{\dagger },
\label{Kraus representation}
\end{equation}%
which satisfy the completeness relationship $\sum_{i}$ $A_{i}^{\dagger
}A_{i}=I_{s}$.

In comparison to classical channels, which are characterized by a unique
capacity \cite{Shannon 1948, Thomas and Cover}, the situation in\ quantum
realm is more rich and complicated. The capacity of a quantum channel
depends on the type of information transmitted, resources shared and
protocols allowed \cite{Bennett and Shor 2004}. In this work, we are
interested in the classical and quantum capacities assisted by entanglement.

\section{Entanglement-assisted capacities}

The capacity of a quantum channel can be enhanced either by encoding
information on entangled states [25-27] or sharing entanglement between
sender and receiver \cite{Bennett and Wiesner 1992}, prior to the
communication. The amount of classical information transmitted across a
quantum channel, provided the sender and receiver share unlimited prior
entanglement is given by the \textit{entanglement-assisted classical
capacity }$C_{E}$ [17-19]. It is obtained by quantum mutual information \cite%
{Nielson and Chuang}, maximized over the input state $\rho _{s}$ i.e.,%
\begin{equation}
C_{E}=\max_{\rho _{s}}\left[ S\left( \rho _{s}\right) +S\left( \mathcal{N}%
\left( \rho _{s}\right) \right) -S_{e}\left( \rho _{s}\right) \right] ,
\label{Entanglement assisted classical capacity}
\end{equation}%
where $S\left( \rho \right) =-$Tr$\left[ \rho \log _{2}\rho \right] $ is the
von Neumann entropy and%
\begin{equation}
S_{e}\left( \rho _{s}\right) =\sum_{i,j}\text{Tr}\left( A_{i}\rho
_{s}A_{j}^{\dagger }\right) \left\vert e_{i}\right\rangle \left\langle
e_{j}\right\vert ,  \label{Entropy exchange}
\end{equation}%
is the entropy exchange \cite{Nielson and Chuang}. Pure-state entanglement
consumed per channel use is $S\left( \rho _{s}\right) $, with $\rho _{s}$
maximizing Eq. (\ref{Entanglement assisted classical capacity}). It is
additive \cite{Holevo 2002}, unlike the classical capacity $C$ \cite%
{Hastings 2009}, and quantum capacity $Q$ \cite{Barnum Nielson Schumacher
1998}, of quantum channels. The \textit{entanglement-assisted quantum
capacity} $Q_{E}=C_{E}/2$ can be determined by superdense coding \cite%
{Bennett and Wiesner 1992} and quantum teleportation \cite{Bennett Brassard
Crepeau Jozsa Peres and Wootters 1993}.

If the entanglement shared prior to the communication is limited then the 
\textit{classical capacity assisted by limited entanglement} of a quantum
channel is given by \cite{Shor 2004},%
\begin{eqnarray}
C_{E}^{\lim } &=&\max_{p_{i},\rho _{s_{i}}}\left[ \sum_{i}p_{i}S\left( \rho
_{s_{i}}\right) +S\left\{ \mathcal{N}\left( \sum_{i}p_{i}\rho
_{s_{i}}\right) \right\} -\sum_{i}p_{i}S_{e}\left( \rho _{s_{i}}\right) %
\right] ,  \notag \\
&&  \label{Classical capacity assisted by limited entanglement}
\end{eqnarray}%
with $\sum_{i}p_{i}S\left( \rho _{s_{i}}\right) \leq P$, where $P$ is the
amount of entanglement available. The maximization is performed over the
input states $\rho _{s_{i}}$ and probability distribution $p_{i}$ with $%
\sum_{i}p_{i}=1$. This provides a trade off curve of the classical capacity
as a function of the amount of entanglement shared. If the shared
entanglement is sufficiently large then Eq. (\ref{Classical capacity
assisted by limited entanglement}) gives $C_{E}$ while for $P=0$ it reduces
to the classical capacity $C$ \cite{Hausladen Jozsa Schumacher 1996,
Schumacher and Westmoreland 1997}.

\section{Time correlated amplitude-damping channel}

Energy dissipation from a quantum system, such as, spontaneous emission of
an atom and relaxation of a spin system at high temperature into the
equilibrium state \cite{Nielson and Chuang}, is modeled by amplitude damping
channel. It is a non-unital channel, i. e., $\mathcal{N}\left( I\right) \neq
I$ \cite{King and Ruskai 2001}, with Kraus operators%
\begin{equation}
A_{0}=\left( 
\begin{array}{cc}
\cos \chi & 0 \\ 
0 & 1%
\end{array}%
\right) ,A_{1}=\left( 
\begin{array}{cc}
0 & 0 \\ 
\sin \chi & 0%
\end{array}%
\right) ,  \label{ADC Kraus}
\end{equation}%
where $\chi $ is the damping parameter with $0\leq \chi \leq \frac{\pi }{2}$%
. We consider an amplitude-damping channel with time-correlated Markov noise
for two consecutive uses \cite{Yeo and Skeen 2003}, given by%
\begin{equation}
\mathcal{N}\left( \rho _{s}\right) =\left( 1-\mu \right)
\sum_{i,j=0}^{1}A_{ij}^{u}\rho _{s}A_{ij}^{u\dag }+\mu
\sum_{k=0}^{1}A_{kk}^{c}\rho _{s}A_{kk}^{c\dag },  \label{Partial ADC}
\end{equation}%
where $0\leq \mu \leq 1$, is the memory parameter. The channel noise is
uncorrelated with probability $\left( 1-\mu \right) $ described by Kraus
operators%
\begin{equation}
A_{ij}^{u}=A_{i}\otimes A_{j},  \label{Un-correlated ADC}
\end{equation}%
where $i,j=0,1$ with $A_{i}$ given by Eq. (\ref{ADC Kraus}), while with
probability $\mu $\ it is correlated and given by $A_{kk}^{c}$. The Kraus
operators $A_{kk}^{c}$ for time-correlated amplitude damping channel for two
channel uses are determined by solving the Lindbladian \cite{Yeo and Skeen
2003}, 
\begin{eqnarray}
\mathcal{L}\rho _{s} &=&-\frac{\alpha }{2}\left[ \left( \sigma ^{\dag
}\otimes \sigma ^{\dag }\right) \left( \sigma \otimes \sigma \right) \rho
_{s}+\rho _{s}\left( \sigma ^{\dag }\otimes \sigma ^{\dag }\right) \left(
\sigma \otimes \sigma \right) \right.  \notag \\
&&\left. -2\left( \sigma \otimes \sigma \right) \rho _{s}\left( \sigma
^{\dag }\otimes \sigma ^{\dag }\right) \right] .
\label{Lindbladian ADC-Two uses}
\end{eqnarray}%
The parameter $\alpha $ is analogous to the Einstein coefficient for
spontaneous emission \cite{Daffer Wodkiewicz McIver 2003}, and $\sigma
^{\dag }\equiv \frac{1}{2}\left( \sigma ^{x}+i\sigma ^{y}\right) $ and $%
\sigma \equiv \frac{1}{2}\left( \sigma ^{x}-i\sigma ^{y}\right) $ are the
raising and lowering operators, respectively. It is solved by using the
damping basis method \cite{Daffer Wodkiewicz McIver 2003, Yeo and Skeen 2003}%
. The map%
\begin{equation}
\Phi \left( \rho \right) =\exp \left( \mathcal{L}t\right) \rho =\sum_{i}%
\text{Tr}\left( L_{i}\rho \right) \exp \left( \lambda _{i}t\right) R_{i},
\label{Damping basis}
\end{equation}%
describes a wide class of Markov quantum channels, where $\lambda _{i}$ are
the damping eigenvalues. The right eigen operators $R_{i}$ satisfy the
eigenvalue equation%
\begin{equation}
\mathcal{L}R_{i}=\lambda _{i}R_{i},  \label{Eigenvalue equation}
\end{equation}%
and the duality relation%
\begin{equation}
\text{Tr}\left( L_{i}R_{i}\right) =\delta _{ij},  \label{Duality relation}
\end{equation}%
with the left eigen operators $L_{i}$. The resulting completely positive and
trace preserving map is given by 
\begin{equation}
\mathcal{E}\left( \rho _{s}\right) =\sum_{i,j=0}^{3}\text{Tr}\left(
L_{ij}\rho _{s}\right) \exp \left( \lambda _{ij}t\right)
R_{ij}=\sum_{k=0}^{1}A_{kk}^{c}\rho _{s}A_{kk}^{c},  \label{Correlated ADC}
\end{equation}%
where the Kraus operators for correlated noise are%
\begin{equation}
A_{00}^{c}=\left( 
\begin{array}{cccc}
\cos \chi & 0 & 0 & 0 \\ 
0 & 1 & 0 & 0 \\ 
0 & 0 & 1 & 0 \\ 
0 & 0 & 0 & 1%
\end{array}%
\right) ,\text{ }A_{11}^{c}=\left( 
\begin{array}{cccc}
0 & 0 & 0 & 0 \\ 
0 & 0 & 0 & 0 \\ 
0 & 0 & 0 & 0 \\ 
\sin \chi & 0 & 0 & 0%
\end{array}%
\right) .  \label{Correlated ADC Kraus}
\end{equation}%
Eqs. (\ref{Partial ADC}), (\ref{Un-correlated ADC}) and (\ref{Correlated ADC
Kraus}) give the time-correlated amplitude-damping channel with memory.

\section{Entanglement-assisted capacities of time-correlated
amplitude-damping channel}

We now calculate the entanglement-assisted capacities of an
amplitude-damping channel with time-correlated Markov noise, for two channel
uses. Consider the protocol where the sender and receiver share two ( same
or different) maximally entangled Bell states \cite{Bell 1966}, prior to the
communication%
\begin{eqnarray}
\left\vert \psi ^{\pm }\right\rangle &=&\frac{1}{\sqrt{2}}\left( \left\vert
00\right\rangle \pm \left\vert 11\right\rangle \right) ,  \notag \\
\left\vert \phi ^{\pm }\right\rangle &=&\frac{1}{\sqrt{2}}\left( \left\vert
01\right\rangle \pm \left\vert 10\right\rangle \right) .  \label{Bell states}
\end{eqnarray}%
The first qubits of the shared states belong to the sender while the second
qubits are in possession of the receiver. The state input to the channel $%
\rho _{s}$\ is given by%
\begin{equation}
\rho _{s}=\text{Tr}_{r}\left( \left\vert \psi ^{+}\right\rangle \left\langle
\psi ^{+}\right\vert \right) \otimes \text{Tr}_{r}\left( \left\vert \phi
^{+}\right\rangle \left\langle \phi ^{+}\right\vert \right) =\frac{I}{4},
\label{Channel input}
\end{equation}%
where $I$ is the $4\times 4$ identity matrix. Information is encoded by the
sender on the input state $\rho _{s}$ and transmitted over the
amplitude-damping channel. The channel maps it to an output state $\mathcal{N%
}\left( \rho _{S}\right) $, given by Eq. (\ref{Partial ADC}) with eigenvalues%
\begin{eqnarray}
\omega _{1} &=&\frac{1}{4}\left[ \left( 1-\mu \right) \cos ^{4}\chi +\mu
\cos ^{2}\chi \right] ,  \notag \\
\omega _{2} &=&\omega _{3}=\frac{1}{4}\left[ \left( 1-\mu \right) \cos
^{2}\chi \left( 2-\cos ^{2}\chi \right) +\mu \right] ,  \notag \\
\omega _{4} &=&-\frac{1}{4}\left( 2-\cos ^{2}\chi \right) \left[ \left(
1-\mu \right) \cos ^{2}\chi +\mu -2\right] .  \label{Output eigenvalues}
\end{eqnarray}%
The amount of information lost due to coupling with the environment during
the transmission is determined by calculating the entropy exchange $%
S_{e}\left( \rho _{s}\right) $ given by Eq. (\ref{Entropy exchange}). We
assume without loss of generality that initially the state of the
environment is pure, 
\begin{equation}
\rho _{e}=\left( \left\vert 00\right\rangle \left\langle 00\right\vert
\right) _{e},  \label{Environment input}
\end{equation}%
which is modified to%
\begin{eqnarray}
\rho _{e}^{\prime } &=&\left( 1-\mu \right) \sum_{i,j,k,l=0}^{1}\text{Tr}%
_{s}\left( A_{ij}^{u}\rho _{s}A_{kl}^{u\dag }\right) \left\vert
e_{ij}\right\rangle \left\langle e_{kl}\right\vert  \notag \\
&&+\mu \sum_{m,n=0}^{1}\text{Tr}_{s}\left( A_{mm}^{c}\rho _{s}A_{nn}^{c\dag
}\right) \left\vert e_{mm}\right\rangle \left\langle e_{nn}\right\vert ,
\label{Environment Output}
\end{eqnarray}%
after interaction with the input state $\rho _{s}$. In the above expression $%
\left\vert e_{ij}\right\rangle =\left\vert e_{i}\right\rangle \otimes
\left\vert e_{j}\right\rangle $ are the orthonormal basis of the environment
and eigenvalues of the output state $\rho _{e}^{\prime }$ are%
\begin{eqnarray}
\widetilde{\omega }_{1} &=&\frac{1}{4}\left[ \left( 1-\mu \right) \sin
^{4}\chi +\mu \sin ^{2}\chi \right] ,  \notag \\
\widetilde{\omega }_{2} &=&\widetilde{\omega }_{3}=\frac{1}{4}\left( 1-\mu
\right) \left( 2-\sin ^{2}\chi \right) \sin ^{2}\chi ,  \notag \\
\widetilde{\omega }_{4} &=&\frac{1}{4}\left[ \left( 1-\mu \right) \left(
2-\sin ^{2}\chi \right) ^{2}+\mu \left( 4-\sin ^{2}\chi \right) \right] .
\label{Environment output eigenvalues}
\end{eqnarray}%
The entanglement-assisted classical capacity for amplitude-damping channel
with correlated noise calculated by using Eq. (\ref{Entanglement assisted
classical capacity}) is%
\begin{equation}
C_{E}^{2}=2-\sum_{i=1}^{4}\left( \omega _{i}\log _{2}\omega _{i}-\widetilde{%
\omega }_{i}\log _{2}\widetilde{\omega }_{i}\right) .  \label{Ce}
\end{equation}

\begin{figure}[tph]
\centering
\includegraphics[width=2.4in]{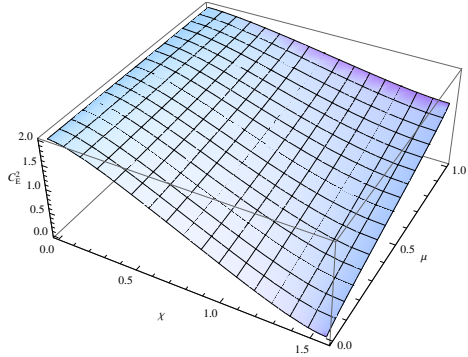}
\caption{Entanglement-assisted classical capacity $C_{E}^{2}$\ as a function
of channel noise $\protect\chi $ and memory $\protect\mu $, normalized with
respect to the number of channel uses. See text for details.}
\label{C_E versus Kie and Meu}
\end{figure}

In Fig. (\ref{C_E versus Kie and Meu}) we plot the capacity $C_{E}^{2}$
against the channel noise $\chi $ and memory parameter $\mu $.
Amplitude-damping channel affects both the populations and coherences of the
quantum system. Therefore, the capacity $C_{E}^{2}$ decreases as the channel
becomes noisy and reduces to zero for maximum value of channel noise.
However, as we increase the memory $\mu $ the capacity $C_{E}^{2}$ increases
for all values of channel noise parameter $\chi $. We infer that memory of
the channel suppresses the noise introduced by it and increases the
capacity. It is interesting to note that for memoryless amplitude-damping
channel the capacity $C_{E}^{2}$ is zero for maximum channel noise $\chi =%
\frac{\pi }{2}$. However, if the channel has finite memory its capacity to
transmit information is non-zero, always. This is similar to the
superactivation phenomenon, where two channel with zero quantum capacity can
transmit quantum information if used together \cite{Smith and Yard 2008}. It
reveals the richness of quantum communication theory where whether a channel
can transmit information depends on the context. The entanglement-assisted
quantum capacity is given by%
\begin{equation}
Q_{E}^{2}=\frac{C_{E}^{2}}{2}=1-\sum_{i=1}^{4}\frac{1}{2}\left( \omega
_{i}\log _{2}\omega _{i}-\widetilde{\omega }_{i}\log _{2}\widetilde{\omega }%
_{i}\right) .  \label{Qe}
\end{equation}

Next we calculate the classical capacity assisted by limited entanglement.
In comparison to the unlimited entanglement shared above an ensemble of
orthogonal states%
\begin{eqnarray}
\left\vert \nu _{1}\right\rangle  &=&\cos \theta _{1}\left\vert
00\right\rangle +\sin \theta _{1}\left\vert 11\right\rangle ,  \notag \\
\left\vert \nu _{2}\right\rangle  &=&\sin \theta _{1}\left\vert
00\right\rangle -\cos \theta _{1}\left\vert 11\right\rangle ,  \notag \\
\left\vert \nu _{3}\right\rangle  &=&\cos \theta _{1}\left\vert
01\right\rangle +\sin \theta _{1}\left\vert 10\right\rangle ,  \notag \\
\left\vert \nu _{4}\right\rangle  &=&\sin \theta _{1}\left\vert
01\right\rangle -\cos \theta _{1}\left\vert 10\right\rangle ,
\label{Ansatz 1}
\end{eqnarray}%
and%
\begin{eqnarray}
\left\vert \upsilon _{1}\right\rangle  &=&\cos \theta _{2}\left\vert
00\right\rangle +\sin \theta _{2}\left\vert 11\right\rangle ,  \notag \\
\left\vert \upsilon _{2}\right\rangle  &=&\sin \theta _{2}\left\vert
00\right\rangle -\cos \theta _{2}\left\vert 11\right\rangle ,  \notag \\
\left\vert \upsilon _{3}\right\rangle  &=&\cos \theta _{2}\left\vert
01\right\rangle +\sin \theta _{2}\left\vert 10\right\rangle ,  \notag \\
\left\vert \upsilon _{4}\right\rangle  &=&\sin \theta _{2}\left\vert
01\right\rangle -\cos \theta _{2}\left\vert 10\right\rangle ,
\label{Ansatz 2}
\end{eqnarray}%
is shared prior to the communication, with $0\leq \theta _{1},\theta
_{2}\leq \frac{\pi }{4}$. The input states $\rho _{s_{i}}$ are 
\begin{equation}
\rho _{s_{i}}=\text{Tr}_{r}\left( \left\vert \nu _{j}\right\rangle
\left\langle \nu _{j}\right\vert \right) \otimes \text{Tr}_{r}\left(
\left\vert \upsilon _{k}\right\rangle \left\langle \upsilon _{k}\right\vert
\right) ,  \label{Limited input}
\end{equation}%
where $j,k=1,\ldots 4$ and%
\begin{eqnarray}
\rho _{s_{1}} &=&\cos ^{2}\theta _{1}\cos ^{2}\theta _{2}\left\vert
00\right\rangle \left\langle 00\right\vert +\cos ^{2}\theta _{1}\sin
^{2}\theta _{2}\left\vert 01\right\rangle \left\langle 01\right\vert   \notag
\\
&&+\sin ^{2}\theta _{1}\cos ^{2}\theta _{2}\left\vert 10\right\rangle
\left\langle 10\right\vert +\sin ^{2}\theta _{1}\sin ^{2}\theta
_{2}\left\vert 11\right\rangle \left\langle 11\right\vert ,  \notag \\
\rho _{s_{2}} &=&\cos ^{2}\theta _{1}\sin ^{2}\theta _{2}\left\vert
00\right\rangle \left\langle 00\right\vert +\cos ^{2}\theta _{1}\cos
^{2}\theta _{2}\left\vert 01\right\rangle \left\langle 01\right\vert   \notag
\\
&&+\sin ^{2}\theta _{1}\sin ^{2}\theta _{2}\left\vert 10\right\rangle
\left\langle 10\right\vert +\sin ^{2}\theta _{1}\cos ^{2}\theta
_{2}\left\vert 11\right\rangle \left\langle 11\right\vert ,  \notag \\
\rho _{s_{3}} &=&\sin ^{2}\theta _{1}\cos ^{2}\theta _{2}\left\vert
00\right\rangle \left\langle 00\right\vert +\sin ^{2}\theta _{1}\sin
^{2}\theta _{2}\left\vert 01\right\rangle \left\langle 01\right\vert   \notag
\\
&&+\cos ^{2}\theta _{1}\cos ^{2}\theta _{2}\left\vert 10\right\rangle
\left\langle 10\right\vert +\cos ^{2}\theta _{1}\sin ^{2}\theta
_{2}\left\vert 11\right\rangle \left\langle 11\right\vert ,  \notag \\
\rho _{s_{4}} &=&\sin ^{2}\theta _{1}\sin ^{2}\theta _{2}\left\vert
00\right\rangle \left\langle 00\right\vert +\sin ^{2}\theta _{1}\cos
^{2}\theta _{2}\left\vert 01\right\rangle \left\langle 01\right\vert   \notag
\\
&&+\cos ^{2}\theta _{1}\sin ^{2}\theta _{2}\left\vert 10\right\rangle
\left\langle 10\right\vert +\cos ^{2}\theta _{1}\cos ^{2}\theta
_{2}\left\vert 11\right\rangle \left\langle 11\right\vert ,
\label{Limited input states}
\end{eqnarray}%
Therefore, for all $\rho _{s_{i}}$%
\begin{equation}
S\left( \rho _{s_{i}}\right) =-\sum_{i=1}^{4}\vartheta _{i}\log
_{2}\vartheta _{i},  \label{Limited input entropy}
\end{equation}%
with%
\begin{eqnarray}
\vartheta _{1} &=&\cos ^{2}\theta _{1}\cos ^{2}\theta _{2},  \notag \\
\vartheta _{2} &=&\cos ^{2}\theta _{1}\sin ^{2}\theta _{2},  \notag \\
\vartheta _{3} &=&\sin ^{2}\theta _{1}\cos ^{2}\theta _{2},  \notag \\
\vartheta _{4} &=&\sin ^{2}\theta _{1}\sin ^{2}\theta _{2}.
\label{Limited input eigenvalues}
\end{eqnarray}%
It has been established that the maximization over input probabilities in
Eq. (\ref{Classical capacity assisted by limited entanglement}) is achieved
when the input states are equiprobable \cite{Arshed Toor Lidar 2010}.
Therefore, $\rho _{s}=\sum_{i}p_{i}\rho _{s_{i}}=\frac{I}{4}$ which after
transmission thorough the channel is mapped to $\mathcal{N}\left( \rho
_{s}\right) $ with eigenvalues given by Eq. (\ref{Output eigenvalues}). Once
again we assume that initially the environment is in a pure state $\rho _{e}$
given by Eq. (\ref{Environment input}) and calculate the entropy exchange
for $\rho _{s_{i}}$ using Eq. (\ref{Environment Output}). The classical
capacity assisted by limited entanglement is calculated using Eq. (\ref%
{Classical capacity assisted by limited entanglement}) as%
\begin{equation}
C_{E}^{\lim }=-\sum_{i=1}^{4}\left[ \vartheta _{i}\log _{2}\vartheta
_{i}+\omega _{i}\log _{2}\omega _{i}\right] +\frac{1}{4}\sum_{i,j=1}^{4}%
\widetilde{\omega }_{s_{i}}^{j}\log _{2}\widetilde{\omega }_{s_{i}}^{j},
\label{Limited Ce for ADC}
\end{equation}%
where%
\begin{eqnarray*}
\widetilde{\omega }_{s_{1}}^{1} &=&\left( 1-\mu \right) \left( \cos
^{2}\theta _{1}\cos ^{2}\chi +\sin ^{2}\theta _{1}\right) \left( \cos
^{2}\theta _{2}\cos ^{2}\chi +\sin ^{2}\theta _{2}\right)  \\
&&+\mu \left[ \sin ^{2}\theta _{1}+\cos ^{2}\theta _{1}\left( \cos
^{2}\theta _{2}\cos ^{2}\chi +\sin ^{2}\theta _{2}\right) \right] , \\
\widetilde{\omega }_{s_{1}}^{2} &=&\left( 1-\mu \right) \cos ^{2}\theta
_{2}\left( \cos ^{2}\theta _{1}\cos ^{2}\chi +\sin ^{2}\theta _{1}\right)
\sin ^{2}\chi , \\
\widetilde{\omega }_{s_{1}}^{3} &=&\left( 1-\mu \right) \cos ^{2}\theta
_{1}\left( \cos ^{2}\theta _{2}\cos ^{2}\chi +\sin ^{2}\theta _{2}\right)
\sin ^{2}\chi , \\
\widetilde{\omega }_{s_{1}}^{4} &=&\cos ^{2}\theta _{1}\cos ^{2}\theta
_{2}\sin ^{2}\chi \left[ \left( 1-\mu \right) \sin ^{2}\chi +\mu \right] , \\
\widetilde{\omega }_{s_{2}}^{1} &=&\left( 1-\mu \right) \left( \cos
^{2}\theta _{1}\cos ^{2}\chi +\sin ^{2}\theta _{1}\right) \left( \cos
^{2}\theta _{2}+\cos ^{2}\chi \sin ^{2}\theta _{2}\right)  \\
&&+\mu \left[ \sin ^{2}\theta _{1}+\cos ^{2}\theta _{1}\left( \cos
^{2}\theta _{2}+\cos ^{2}\chi \sin ^{2}\theta _{2}\right) \right] , \\
\widetilde{\omega }_{s_{2}}^{2} &=&\left( 1-\mu \right) \sin ^{2}\theta
_{2}\left( \cos ^{2}\theta _{1}\cos ^{2}\chi +\sin ^{2}\theta _{1}\right)
\sin ^{2}\chi , \\
\widetilde{\omega }_{s_{2}}^{3} &=&\left( 1-\mu \right) \cos ^{2}\theta
_{1}\left( \cos ^{2}\theta _{2}+\cos ^{2}\chi \sin ^{2}\theta _{2}\right)
\sin ^{2}\chi , \\
\widetilde{\omega }_{s_{2}}^{4} &=&\cos ^{2}\theta _{1}\sin ^{2}\theta
_{2}\sin ^{2}\chi \left[ \left( 1-\mu \right) \sin ^{2}\chi +\mu \right] , \\
\widetilde{\omega }_{s_{3}}^{1} &=&\left( 1-\mu \right) \left( \cos
^{2}\theta _{1}+\cos ^{2}\chi \sin ^{2}\theta _{1}\right) \left( \cos
^{2}\theta _{2}\cos ^{2}\chi +\sin ^{2}\theta _{2}\right)  \\
&&+\mu \left[ \cos ^{2}\theta _{1}+\sin ^{2}\theta _{1}\left( \cos
^{2}\theta _{2}\cos ^{2}\chi +\sin ^{2}\theta _{2}\right) \right] , \\
\widetilde{\omega }_{s_{3}}^{2} &=&\left( 1-\mu \right) \cos ^{2}\theta
_{2}\left( \cos ^{2}\theta _{1}+\cos ^{2}\chi \sin ^{2}\theta _{1}\right)
\sin ^{2}\chi , \\
\widetilde{\omega }_{s_{3}}^{3} &=&\left( 1-\mu \right) \sin ^{2}\theta
_{1}\left( \cos ^{2}\theta _{2}\cos ^{2}\chi +\sin ^{2}\theta _{2}\right)
\sin ^{2}\chi , \\
\widetilde{\omega }_{s_{3}}^{4} &=&\cos ^{2}\theta _{2}\sin ^{2}\theta
_{1}\sin ^{2}\chi \left[ \left( 1-\mu \right) \sin ^{2}\chi +\mu \right] ,
\end{eqnarray*}%
\begin{eqnarray}
\widetilde{\omega }_{s_{4}}^{1} &=&\left( 1-\mu \right) \left( \cos
^{2}\theta _{1}+\cos ^{2}\chi \sin ^{2}\theta _{1}\right) \left( \cos
^{2}\theta _{2}+\cos ^{2}\chi \sin ^{2}\theta _{2}\right)   \notag \\
&&+\mu \left[ \cos ^{2}\theta _{1}+\sin ^{2}\theta _{1}\left( \cos
^{2}\theta _{2}+\cos ^{2}\chi \sin ^{2}\theta _{2}\right) \right] ,  \notag
\\
\widetilde{\omega }_{s_{4}}^{2} &=&\left( 1-\mu \right) \sin ^{2}\theta
_{2}\left( \cos ^{2}\theta _{1}+\cos ^{2}\chi \sin ^{2}\theta _{1}\right)
\sin ^{2}\chi ,  \notag \\
\widetilde{\omega }_{s_{4}}^{3} &=&\left( 1-\mu \right) \sin ^{2}\theta
_{1}\left( \cos ^{2}\theta _{2}+\cos ^{2}\chi \sin ^{2}\theta _{2}\right)
\sin ^{2}\chi ,  \notag \\
\widetilde{\omega }_{s_{4}}^{4} &=&\sin ^{2}\theta _{1}\sin ^{2}\theta
_{2}\sin ^{2}\chi \left[ \left( 1-\mu \right) \sin ^{2}\chi +\mu \right] .
\label{Limited entropy exchange eigenvalues}
\end{eqnarray}%
The capacity $C_{E}^{\lim }$ increases with the increase in $\theta _{1}$
and $\theta _{2}$ which results into higher amount of shared entanglement.
It acquires the maximum value for $\theta _{1}=\theta _{2}=\frac{\pi }{4}$,
for which the shared states are maximally entangled and Eq. (\ref{Limited Ce
for ADC}) reduces to Eq. (\ref{Ce}). For $\theta _{1}=\theta _{2}=0$, Eq. (%
\ref{Limited Ce for ADC}) reduces to the product state classical capacity $%
C_{p}$\ for time-correlated amplitude-damping channel given as%
\begin{equation}
C_{p}^{2}=-\sum_{i=1}^{4}\omega _{i}\log _{2}\omega _{i}+\frac{1}{4}%
\sum_{i,j=1}^{4}\widetilde{\omega }_{s_{i}}^{j}\log _{2}\widetilde{\omega }%
_{s_{i}}^{j},  \label{Product state classical capacity}
\end{equation}%
with%
\begin{eqnarray}
\widetilde{\omega }_{s_{1}}^{1} &=&\left( 1-\mu \right) \cos ^{4}\chi +\mu
\cos ^{2}\chi ,  \notag \\
\widetilde{\omega }_{s_{1}}^{2} &=&\widetilde{\omega }_{s_{1}}^{3}=\left(
1-\mu \right) \cos ^{2}\chi \sin ^{2}\chi ,  \notag \\
\widetilde{\omega }_{s_{1}}^{4} &=&\left( 1-\mu \right) \sin ^{4}\chi +\mu
\sin ^{2}\chi ,  \notag \\
\widetilde{\omega }_{s_{2}}^{1} &=&\left( 1-\mu \right) \cos ^{2}\chi +\mu ,
\notag \\
\widetilde{\omega }_{s_{2}}^{2} &=&0,  \notag \\
\widetilde{\omega }_{s_{2}}^{3} &=&\left( 1-\mu \right) \sin ^{2}\chi , 
\notag \\
\widetilde{\omega }_{s_{2}}^{4} &=&0,  \notag \\
\widetilde{\omega }_{s_{3}}^{1} &=&\left( 1-\mu \right) \cos ^{2}\chi +\mu ,
\notag \\
\widetilde{\omega }_{s_{3}}^{2} &=&\left( 1-\mu \right) \sin ^{2}\chi , 
\notag \\
\widetilde{\omega }_{s_{3}}^{3} &=&\widetilde{\omega }_{s_{3}}^{4}=0,  \notag
\\
\widetilde{\omega }_{s_{4}}^{1} &=&1,  \notag \\
\widetilde{\omega }_{s_{4}}^{2} &=&\widetilde{\omega }_{s_{4}}^{3}=%
\widetilde{\omega }_{s_{4}}^{4}=0.  \label{Classical capacity eigenvalues}
\end{eqnarray}

\begin{figure}[tph]
\centering
\includegraphics[width=2.4in]{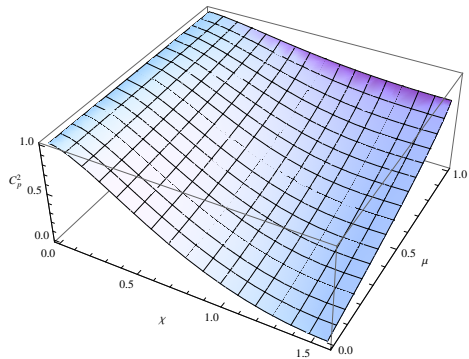}
\caption{Classical capacity $C_{p}^{2}$ for product states as a function of
channel noise $\protect\chi $ and memory $\protect\mu $, normalized with
respect to the number of channel uses.}
\label{C_p versus Kie and Meu}
\end{figure}

We give the plot of capacity $C_{p}^{2}$ as function of channel noise $\chi $
and memory $\mu $ in Fig. (\ref{C_p versus Kie and Meu}). It is evident that
the capacity decreases as the channel becomes noisy, reducing to zero for
maximum channel noise $\chi =\frac{\pi }{2}$. In the presence of memory, the
capacity $C_{p}^{2}$ is non-zero independent of channel noise $\chi $ and
increases with the channel memory attaining maximum value for perfect memory
channel i. e., $\mu =1$. This is in agreement with the numerical study for
product state classical capacity of amplitude-damping channel presented in 
\cite{Yeo and Skeen 2003}.

\section{Conclusion}

We have studied an amplitude-damping channel for transmission of
information, provided entanglement is shared prior to the communication. The
noise over consecutive uses of the channel is time-correlated Markov noise.
We analytically determine the classical and quantum capacities for this
channel which exhibit strong dependence on channel memory. The capacities
decrease as the channel noise increases reducing to zero for maximum channel
noise. However, if the channel has memory its capacity to transmit
information is always non-zero. Memory increases the predictability of the
channel action on the successive input states and thus suppresses the noise
introduced by the channel. In the presence of channel memory, its capacity
increases acquiring maximum value for perfect memory channel. We also
calculate the classical capacity assisted by limited entanglement which
increases with the amount of shared entanglement. It gives
entanglement-assisted classical capacity if maximally entangled states are
shared while reduces to product state classical capacity when the amount of
shared entanglement reduces to zero. For a given value of entanglement, the
classical capacity assisted by limited entanglement increases with the
channel memory, independent of channel noise. It is always non-zero in the
presence of channel memory attaining the maximum value for perfect memory.

\begin{acknowledgement}
N. Arshed was supported by the Higher Education Commission Pakistan under
Grant No. 063-111368-Ps3-001.
\end{acknowledgement}

\end{document}